\documentclass[11pt]{article}

\usepackage{amssymb,amsmath}
\usepackage[dvips]{graphicx}
\usepackage{epsfig}
\usepackage{cite}

\def\laq{~\raise 0.4ex\hbox{$<$}\kern -0.8em\lower 0.62
ex\hbox{$\sim$}~}
\def\gaq{~\raise 0.4ex\hbox{$>$}\kern -0.7em\lower 0.62
ex\hbox{$\sim$}~}
\numberwithin{equation}{section}

\def\be{\begin{equation}}
\def\ee{\end{equation}}
\def\bea{\begin{eqnarray}}
\def\eea{\end{eqnarray}}
\newcommand{\nn}{\nonumber}
\newcommand{\de}{\partial}

\def \Mp {M_{\rm P}}

\def \ra {\rightarrow}
\def \l {\lambda}
\def \L {\Lambda}
\def \d {\delta}

\def \b {\beta}
\def \a {\alpha}
\def \b {\beta}

\def \g {\gamma}
\def \s {\sigma}

\def \e {\epsilon}

\def \om {\omega}
\def \Om {\Omega}

\def \5 {{}^{(5)}}
\def \psih {\hat \psi}
\def \dag {\dagger}
\def \omk {\omega_{m,k}}
\def \lb {\bar{\lambda}}
\def \mb {\bar{\mu}}

\begin{document}

\begin{titlepage}

\begin{flushright}
BA-TH/06-530 \\
CERN-PH-TH/2006-096 \\
gr-qc/0601132
\end{flushright}

\vspace*{1.5 cm}

\begin{center}

\Huge
{Massless and massive graviton spectra \\ in anisotropic dilatonic \\ brane-world cosmologies}

\vspace{1cm}

\large{G. De Risi}

\normalsize
\vspace{.2in}
{\sl Dipartimento di Fisica, Universit\`a di Bari, \\
and} \\
{\sl Istituto Nazionale di Fisica Nucleare, Sezione di Bari\\
Via G. Amendola 173, 70126 Bari, Italy}

\vspace*{1.5cm}

\begin{abstract}

We consider a brane-world model in which an anisotropic brane is embedded in a dilatonic background.
We find the background solutions and study the behaviour of the perturbations when the Universe
evolves from an inflationary Kasner phase to a Minkowski phase.
We calculate the massless mode spectrum, and find that it does not differ from that expected in
standard four-dimensional cosmological models. We then evaluate the spectrum of both light
(ultrarelativistic) and heavy (nonrelativistic) massive modes, and find that, at high energies,
there can be a strong enhancement of the Kaluza--Klein spectral amplitude, which
can become dominant in the total spectrum. The presence of the dilaton,
on the contrary, decreases the relative importance of the massive modes.

\end{abstract}
\end{center}

\end{titlepage}
\newpage

\section{Introduction}

The brane-world scenario, developed starting from the fundamental work by Randall and Sundrum
\cite{Randall:1999vf}, has received enormous attention in the past years, mostly because it is possible
to generalize it to have interesting cosmological models (see \cite{Maartens:2003tw} for
a review). The fundamental question hence is how these models can be tested, and if there are signals that
could allow us to distinguish between a brane-world model and a more conventional
four-dimensional cosmological one.
Studying the production of cosmological perturbations during inflation is a powerful tool
in trying to answer to this question. Indeed much effort was devoted in building models
(see, for example \cite{Deruelle:2000yj,Koyama:2000cc,Langlois:2000ns,Gorbunov:2001ge,Kobayashi:2003cn,
Easther:2003re,Cavaglia:2005id,Cartier:2005br,Kobayashi:2005dd}) so as to solve the perturbation dynamics,
which is in general very hard to deal with.

In this work we develop a gravi-dilaton brane-world with a non-trivial dynamics on the brane
(a lot of papers have been written generalizing the standard RS solution to a more general framework
in which a scalar field is included, see for example
\cite{Maartens:1999hf,Maeda:2000wr,Bozza:2001xt,Koyama:2003yz,Koyama:2003sb,Kobayashi:2003cb}.
In particular, following our previous work \cite{Cavaglia:2005id},
we consider a higher-dimensional $p$-brane (with $p>3$) coupled with a bulk dilaton. If we do not
put matter on the brane, it is possible to solve exactly the Einstein equations to obtain a Kasner
solution, with $d$ expanding and $n$ contracting dimensions, on the brane itself. The bulk equation
decouples, and is solved by a warp AdS-like factor, while the dilaton grows logarithmically
as it goes away from the brane.

Turning to study the dynamics of the tensor perturbations, we find that
massless and massive modes can be treated independently, and do not mix. We outline a procedure
that allows us to study the production of the gravitational and of the Kaluza--Klein fluctuations;
we also calculate the spectral distribution of these fluctuations, amplified during an imaginary
phase transition from the inflationary Kasner regime to a simple Minkowski phase. This is done
for the massless mode, and for both light and heavy-massive modes with respect to the curvature
scale at the transition epoch. Our results show, differently from what happens in other
models present in the literature, that if the transition occurs at high enough energy, there can be a
strong enhancement of the contribution of the KK modes; this should be in principle
observable, since, in this regime, the complete spectrum is found to be quite different from what
would be expected if there were only the conventional 4D graviton. We also find that, on the contrary,
the dilaton effect results in lowering the amplitude of the KK perturbations.

The paper is organized as follows: in section \ref{backgroundsect} we present and solve the background
equations. In section \ref{perturbations} we study the perturbation around this background, and
define the action that controls the dynamics of each single mode (massless and massive) on the brane.
Section \ref{Actsection} outlines how to develop the canonical analysis of the mode action, and how
to obtain the correct canonical equation that describes the evolution of the mode during
the phase transition. In section \ref{specder} we obtain the correct expression for the spectral
amplitude, which is specialized for the graviton in section \ref{masslesssect}, and for the
KK modes in section \ref{KKsect}. Finally, in section \ref{conclsec}, we make some considerations
on the results obtained, and draw our conclusions.

\section{Background solutions}
\label{backgroundsect}

Let us consider a model in which a brane is non-minimally coupled to a bulk dilaton $\phi$.
We work in a $D$-dimensional space ($D=p+2$), and set the brane fixed at the origin: $X^{D-1}=z=0$. The action is
\bea
S &=& -\frac{M^p}{2} \int d^D x \sqrt{\left| g \right|} \left( R + 2 \L_D e^{\s_1 \phi}
-g^{AB} \de_A \phi \de_B \phi \right) \nn \\
& & -  T_p \int d^{D-1} \xi \sqrt{\left| \g \right|} e^{(p+1)\s_2 \phi},
\label{Action}
\eea
where $\g_{AB} = g_{AB} + n_A n_B$ is the induced metric on the brane,
$T_p$ is the brane tension, and $\L_D$ is the bulk cosmological constant.

Variation of this action with respect to the metric and to the dilaton gives the Einstein equations:
\be
G_{AB} = \left[ \L_D e^{\s_1 \phi} - \frac{1}{2} \left( \de_c \phi \right)^2 \right] g_{AB} +
\de_A \phi \de_B \phi + \sqrt{\frac{|\g|}{|g|}} \frac{T_p}{M^p} e^{(p+1)\s_2 \phi} \g_{AB}~\d(z),
\label{einsteq}
\ee
and the dilaton equation:
\be
\de_A \de^A \phi + \s_1 \L_D e^{\s_1 \phi} + (p+1) \s_2
\sqrt{\frac{|\g|}{|g|}} \frac{T_p}{M^p} e^{(p+1)\s_2 \phi} ~\d(z).
\label{dileq}
\ee
These equations, written as above, already include the Israel junction conditions \cite{Israel:1966rt},
which can be deduced by integrating along a small interval across the brane on a geodesic
perpendicular to the brane itself \cite{Shiromizu:1999wj,Maartens:2003tw} (see also \cite{Aliev:2004ds}).
They read, for the Einstein equations (\ref{einsteq})
\be
\left[ k_{AB} \right] = -\frac{1}{p}\frac{T_p}{M^p} \left(e^{(p+1)\s_2\phi}\g_{AB} \right),
\label{Israel}
\ee
where the square brackets denote the difference between the left and the right limiting value on the brane:
\be
\left[ f \right]= \left( \lim_{\e \ra 0^+} - \lim_{\e \ra 0^-}\right)f(\e),
\label{juncon}
\ee
and $k_{AB}$ is the extrinsic curvature on the brane.

To solve the equations we set the following ansatz on the metric:
\be
ds^2 = f^2(z) \left( dt^2 - a^2(t) {\bf dx^2} - b^2(t) {\bf dy^2} - dz^2 \right),
\label{metricansatz}
\ee
i.e. we allow the metric to be anisotropic, and impose that $d$ spatial dimensions expand and the other $n$
contract $(p=d+n)$. We also impose that the dilaton depend only on the extra-dimension:
$\phi=\phi(z)$. The equations specialize in:
\bea
&& -p F' -\frac{p(p-1)}{2} F^2 + \frac{d(d-1)1}{2}H^2 + \frac{n(n-1)1}{2}G^2 + dn HG \nn \\
&& = \frac{1}{2} \phi'^2 + \Lambda_D e^{\s_1 \phi} f^2
\label{Einst00} \\
&& - p F' - \frac{p(p-1)}{2} F^2  + (d-1) \dot H + \frac{d(d-1)}{2} H^2 + n \dot G +
\frac{n(n+1)}{2}G^2 +(d-1)nHG \nn \\
&& = \frac{1}{2} \phi'^2 + \Lambda_D e^{\s_1 \phi} f^2
\label{Einstij} \\
&& -p F' -\frac{p(p-1)}{2} F^2 + d \dot H + \frac{d(d+1)}{2}H^2 + (n-1) \dot G
+ \frac{n(n-1)}{2}G^2 + d(n-1)H G \nn \\
&& = \frac{1}{2} \phi'^2 + \Lambda_D e^{\s_1 \phi} f^2
\label{Einstab} \\
&& -\frac{p(p+1)}{2}F^2 + d \left(\dot H + \frac{d+1}{2}H^2 \right) + n\left(\dot G
+ \frac{n+1}{2}G^2 \right)
+dn HG \nn \\
&& = -\frac{1}{2} \phi'^2 + \Lambda_D e^{\s_1 \phi} f^2
\label{Einst55} \\
&& \phi'' +pF\phi' - \s_1 \Lambda_D e^{\s_1 \phi} f^2 =0,
\label{dilaton}
\eea
where dots (primes) denote differentiation w.r.t.\ $t$ ($z$), and $H= \dot a/a$, $G= \dot b /b$, $F=f'/f$.
Note that the singular part is absent from the above equations, because it will be taken into account
by satisfying the Israel junction conditions. Inserting the ansatz (\ref{metricansatz}) into
(\ref{Israel}) we get the metric junction condition:
\be
\left[f'\right] = - \frac{1}{p} \frac{T_p}{M^p}
\left. \left(e^{(p+1)\s_2 \phi} f^2 \right) \right|_{z=0}.
\label{Israelmetric}
\ee
The dilaton junction condition is obtained directly by applying on eq. (\ref{dileq})
the integration procedure described above, as in standard one-dimensional quantum mechanics,
and it reads
\be
\left[\phi'\right] = (p+1)\s_2 \frac{T_p}{M^p} \left. \left(e^{(p+1)\s_2 \phi} f \right) \right|_{z=0}.
\label{Israeldilat}
\ee

The Einstein equations (\ref{Einst00}--\ref{Einst55}) can be decoupled, and the time
dependence is solved, as in \cite{Cavaglia:2005id}, by the Kasner solution:
\bea
a(t) = \left|\frac{t}{t_0}\right|^{\l}\, &~~~~~~& b(t) = \left|\frac{t}{t_0}\right|^{\mu}\ \nn \\
\l = \frac{1 \pm  \sqrt{\frac{n(d+n-1)}{d}}}{d+n}\, &~~~~~~& \mu =
\frac{1 \mp  \sqrt{\frac{d(d+n-1)}{n}}}{d+n},
\label{tsol}
\eea
which can describe a superinflationary solution \cite{Gasperini:1992em,Gasperini:2002bn}
on the negative branch of the time axis, if we choose the minus sign in $\l$ and allow
for 2 or more internal dimensions (of course, in the framework of superstring theory there is
room for up to 5 internal compact dimensions if we want 3 external and one warped).
So we are left with the $z$ dependent part of the Einstein equations, which can be rearranged as:
\bea
-pF'- p^2F^2 &=& 2\L_D e^{\s_1 \phi} f^2 \nn \\
-pF'+ pF^2 &=& \phi'^2,
\label{Einstz}
\eea
and the dilaton equation (\ref{dilaton}), which nevertheless depends on the other equations
by means of the Bianchi identities, as expected.

To solve eqs. (\ref{Einstz}) we seek for a solution of the form \cite{Cvetic:2000pn,Bozza:2001xt}:
\bea
f(z) &=& \left( 1+ \frac{z}{z_0} \right)^\a \nn \\
\phi(z) &=& \log \left( 1+ \frac{z}{z_0} \right)^\b;
\label{solCLP}
\eea
here $z_0$ is a positive constant (which corresponds to the AdS length in the usual RS model),
and the solutions are intended to be in the $z>0$ region, since we consider a $Z_2$ symmetric background.
By inserting these expressions in (\ref{Einstz}) and taking into account the Israel junction
conditions (\ref{Israelmetric}) and (\ref{Israeldilat}) we get the following expressions for the exponents
\be
\a =\frac{4}{p\s_1^2-4},~~~~~~\b =-\frac{2p\s_1}{p\s_1^2-4},
\label{zexponents}
\ee
and the following relations between the parameters
\bea
\s_2 &=& \frac{\s_1}{2(p+1)} \nn \\
\frac{T_p}{M^p} &=& \frac{8p}{z_0(4-p\s_1^2)} \nn \\
\L_D &=& -2 \frac{4p(p+1) - p^2\s_1^2}{(z_0(4-p\s_1^2))^2}.
\label{constants}
\eea
To get a positive brane tension and a negative bulk cosmological constant, we assume that:
\be
-\frac{2}{\sqrt{p}}<\s_1<\frac{2}{\sqrt{p}}.
\label{diseq}
\ee
From (\ref{constants}) it is easy to get the relation between the brane tension and the bulk cosmological
constant
\be
\frac{T_p}{M^p}=4\sqrt{\frac{-2\L_D}{4\frac{p+1}{p}-\s_1^2}},
\label{tensvscc}
\ee
which is the generalization of the fine-tuning relation in the standard RS scenario.

It is possible to obtain a different class of solutions, by imposing a different ansatz:
\bea
f(z) &=& e^{-\a z/z_0} \nn \\
\phi(z) &=& -\b \frac{z}{z_0}.
\label{ansatz2}
\eea
This solution saturates the bound (\ref{diseq}) $\s_1=2/\sqrt{p}$, and has the exponents related
as follows
\be
\b = \pm \sqrt{p}\a.
\ee
The relation between the tension and the cosmological constant is unchanged
\be
\frac{T_p}{M^p}=\sqrt{-8\L_D}.
\ee
Nevertheless, we will only consider in what follows the first kind of solution described above.

\section{Perturbation equations}
\label{perturbations}

In this section we are going to derive the equations of motion for the tensor perturbations of the metric,
$\d^{(1)} g_{AB}=h_{AB}$. We set the dilaton perturbation equal to zero, $\d^{(1)} \phi=0$, because it would
decouple from the tensor fluctuations, impose that the perturbation depends on only the external spatial
dimensions and work in the transverse-traceless gauge:
\be
h_{0A}=h_{aA}=h_{D-1,A}=0,~~~~~~g^{ij}h_{ij}=\nabla_jh_i^{~j}=0.
\label{constraints}
\ee
The second-order perturbation of the countervariant indices of the metric and of the
dilaton is set to zero as well. To obtain the expression of the induced metric, we use the equation
\be
g_{AB}n^A n^B = -1 = \left(g^{(0)}_{AB} + h_{AB} \right)
\left(n^{(0)A}+ \d^{(1)} n^A \right) \left(n^{(0)B}+ \d^{(1)} n^B \right).
\label{pertnorm}
\ee
The zeroth-order part of this expression is equal to $-1$, and we can ignore orders
higher than the first. Then eq. (\ref{pertnorm}) becomes:
\be
2g_{AB}n^{(0)A}\d^{(1)} n^B=0=-\frac{1}{f}\d^{(1)} n^5.
\label{normvalue}
\ee
So the extra-dimensional part of the normal unit vector vanishes, and therefore the
complete vector is left unchanged as well, $\d^{(1)} n^A=0$. This shows that
the induced perturbed metric is:
\be
\d^{(1)} \g_{AB} \equiv \bar h_{AB} = \left. h_{AB}\right|_{z=0}.
\label{pertindmetric}
\ee

The perturbation equations can be obtained by variating the action (\ref{Action}) perturbed to order $h^2$.
After some algebraic manipulation and making use of the background equations (\ref{Einst00}--\ref{Einst55}),
we find the final form:
\be
\d^{(2)} S_{(a)} = \frac{M^d}{4} \int d^{d+1}x dz~a^d b^n f^p \left[
\dot h_{(a)}^2 - \sum_{k=1}^d \frac{(\de_k h_{(a)})^2}{a^2} -h_{(a)}^{\prime 2}\right],
\label{pertaction}
\ee
where we have assumed that the internal dimensions have been compactified on a compact manifold of size
$M^{-n}$. This is the action for the single polarization mode: $\d^{(2)} S = \sum_{(a)} \d^{(2)} S_{(a)}$.
The polarization mode $h_{(a)}$ is defined via the spin-2 polarization tensors,
$h_{ij}=h_{(a)} \e^{(a)}_{ij}$, which satisfy the relation: $\e_{~k}^{l~(a)}\e_{~l}^{k~(b)}=2\d^{ab}$.
From now on we will omit the polarization index $a$.
Variation of (\ref{pertaction}) leads to the equation of motion for each mode of the tensor perturbations,
which is, as expected, the $D$-dimensional covariant d'Alembert operator on the background considered:
\be
\ddot h +\left(dH +n G\right) \dot h -\frac{\nabla^2}{a^2}h - h'' - pFh' = 0.
\label{perteq}
\ee

We now need a prescription to project the perturbation equations, which is free of singularities,
on the brane. This condition can be obtained by perturbing to the first order the Israel
junction condition (\ref{Israel}). Making use of the particular form of the metric
(\ref{metricansatz}) we derive the condition:
\be
\left. \left( - \frac{1}{p} \frac{T_p}{M^p}e^{(p+1)\s_2\phi} h \right) \right|_{z=0} =
\left. \left( \frac{1}{f} \left( h' +2F h \right) \right) \right|_{z=0}.
\label{Israelpert}
\ee

To solve the perturbation equation (\ref{perteq}), we can expand its
solutions on an orthogonal basis of functions of $z$ (the precise form of the scalar product, and
consequently of the normalization of the functions, will be discussed shortly):
\be
h(t,x^i,z)= v_0(t,x^i) \psi_0(z) + \int dm~v_m(t, x^i) \psi_m(z) = h_0 + \int dm~h_m.
\label{pertexp}
\ee
Inserting this expansion in (\ref{perteq}) and in (\ref{Israelpert}) leads respectively to the two equations
(which are valid for both the massless and the massive modes):
\bea
\psi''_m +pF\psi'_m = - m^2 \psi_m \label{pertsep1} \\
\ddot v_m +\left( dH +nG \right) \dot v_m -\frac{\nabla^2}{a^2} v_m = - m^2 v_m,
\label{pertsep2}
\eea
and to the condition
\be
\int dm~v_m \left.\left[\frac{1}{f} \left(\psi'_m +2 F \psi_m \right) + \frac{1}{p}\frac{T_p}{M^p}
e^{(p+1)\s_2\phi} \psi_m \right]\right|_{z=0} =0.
\label{Israelexp}
\ee

Let us now substitute the decomposition (\ref{pertexp}) into the action (\ref{pertaction}).
First we consider the zero mode. After integrating by parts and making use of (\ref{pertsep1})
we get
\be
\d^{(2)} S_0 = \frac{M^d}{4} \int dz~f^p |\psi_0|^2 \int d^{d+1}x~a^d b^n \left[
\dot{v}_0^2 - \sum_{i=1}^d \frac{(\de_i v_0)^2}{a^2}\right].
\label{0modeactprov}
\ee
So it is clear that the auxiliary field $\psih_0 = \sqrt{M} f^{\frac{p}{2}} \psi_0$ has the
correct canonical dimension to be normalized to unity:
\be
\int dz~|\psih_0(z)|^2 = M\int dz~f^p |\psi_0(z)|^2 = 1.
\label{zeronormcond}
\ee
From this we can write down the final form for the action that describes the evolution of the zero mode
\bea
\d^{(2)} S_0 &=& \frac{M^{d-1}}{4} \int d^{d+1}x~a^d(t) b^n(t) \left[
\dot{v}_0^2 - \sum_{i=1}^d \frac{(\de_i v_0)^2}{a^2} \right] \nn \\
&=& \frac{M^{d-1}}{4|\psi_0(0)|^2} \int d^{d+1}x~a^d(t) b^n(t) \left[
\dot{\bar h}_0^2 - \sum_{i=1}^d \frac{(\de_i \bar h_0)^2}{a^2} \right],
\label{0pertaction}
\eea
and the effective Planck mass of the 4-dimensional effective graviton can be easily read off to be:
\be
\Mp^{d-1}=\frac{M^{d-1}}{4|\psi_0(0)|^2},
\label{4DPlankmass}
\ee
as in the ordinary RS scenario \cite{Randall:1999vf,Cavaglia:2005id}.

Now we turn to study the massive modes. Again, after a little algebraic manipulation, we arrive at
\bea
\d^{(2)} S &=& \d^{(2)} S_0 + \frac{M^d}{4} \int dm~dm' \int dz~f^p \psi_m \psi_{m'} \nn \\
&& \times \int d^{d+1}x~a^d b^n \left[\dot{v}_m\dot{v}_{m'} -
\sum_{i=1}^d \frac{(\de_i v_m \de_i v_{m'})}{a^2}
- m^2 v_m v_{m'}\right].
\label{massactprov}
\eea
In this case the auxiliary fields with the correct canonical dimension are
$\psih_m = f^{\frac{p}{2}} \psi_m$, which satisfy a Schr\"{o}dinger-like equation
\be
\psih''_m + \left[m^2 - \left(\frac{p^2}{4} F^2 + \frac{p}{2} F'\right) \right] \psih_m = 0;
\label{auxperteq}
\ee
therefore, they can be normalized as a complete set of orthonormal solutions
\be
\int_{-\infty}^{+\infty} dz~ \psih_m \psih_{m'} = \int dz~f^p \psi_m \psi_{m'} = \d(m-m').
\label{norm1}
\ee
By making use of this relation in (\ref{massactprov}) we finally find that the action (\ref{pertaction})
can be thought of as the sum of the massless mode action with an infinite set of massive actions,
one for each single massive mode:
\be
\d^{(2)} S = \d^{(2)} S_0 + \int dm~\d^{(2)} S_m.
\ee
The massless mode action $\d^{(2)} S_0$ of (\ref{0pertaction})
can be interpreted as the 4-dimensional graviton, while the form of the massive mode action is:
\be
\d^{(2)} S_m = \frac{M^d}{4|\psi_m(0)|^2} \int d^{d+1}x~a^d(t) b^n(t) \left[
\dot{\bar h}_m^2 - \sum_{i=1}^d \frac{(\de_i \bar h_m)^2}{a^2} -m^2 \bar h_m^2\right].
\label{4Dpertaction}
\ee
From (\ref{4Dpertaction}) we can deduce the value of the generalized ``Planck mass''
for each mode as the coefficient that multiplies the mode action:
\be
M^d_m = \frac{M^d}{4|\psi_m(0)|^2}.
\label{4Dmasscoupling}
\ee

At this stage, two remarks are in order. The first is that the massive mode action does not
contain terms that couple the massive modes to each other, i.e. each KK mode is ``free'' with respect
to the interaction with the others. This is due to the fact that the perturbation equation (\ref{perteq}) decouples
(at least in the first-order approximation), so that we can treat the two equations (\ref{pertsep2}) independently,
and the Israel junction condition (\ref{Israelexp}) can be satisfied for each massive mode independently,
as we will do in section \ref{KKsect} (this is not true in general, see for example \cite{Cartier:2005br}
in which a similar method leads to an action for the modes that contains coupling terms).
Moreover, it is worth stressing that the massive mode action is not adimensional,
since it has the dimension of a length, $[\d^{(2)}S_m]=[M^{-1}]$. This is of course a consequence
of the fact that the spectrum of the KK modes is continuous, and indicates
that physical quantities will be obtained by integrating over a suitable
mass interval. But, formally, for what concerns the manipulation to obtain the massive mode spectrum,
we will treat (\ref{4Dpertaction}) as a genuine canonical action. Nevertheless,
the important result here is that modes do not mix and can be treated separately.

\section{Canonical analysis of the perturbed action}
\label{Actsection}

To calculate the spectrum of the fluctuation, we need to put the actions (\ref{0pertaction})
and (\ref{4Dpertaction}) in a canonical form. In order to do this we will
adopt the conformal time to describe the evolution of the system (from now on a
dot will denote a derivation with respect to $\eta$):
\be
d \eta = \frac{dt}{a(t)},
\label{diffconf}
\ee
and introduce the pump field
\be
\xi_0(\eta) = \sqrt{\frac{M^{d-1}}{2}}\frac{1}{\psi_0(0)} a^{\frac{d-1}{2}}b^{\frac{n}{2}},~~~~
\xi_m(\eta) = \sqrt{\frac{M^d}{2}}\frac{1}{\psi_m(0)} a^{\frac{d-1}{2}}b^{\frac{n}{2}}.
\label{pumpfield}
\ee
Through the pump field it is possible to introduce the canonical field\footnote{We stress again that
the canonical field for the massive modes actually does not have a canonical dimension.}:
\be
u_m(\eta,x^i) = \xi_m(\eta)\bar h_m(\eta,x^i).
\label{canvar}
\ee
In terms of this canonical field, the action can be expressed in a normal form (valid both for
$m=0$ and for $m\neq0$):
\be
\d^{(2)} S_m = \frac{1}{2} \int d^dx d\eta~\left[
\dot{u}_m^2 - \sum_{i=1}^d (\de_i u_m)^2+\left(\frac{\ddot{\xi}_m}{\xi_m} - m^2a^2\right) u_m^2\right].
\label{4Dconfaction}
\ee
This form for the action makes it possible to adopt the standard quantization procedure: we promote
$u_m$ to operators and impose the canonical commutation relations among the fields
and their conjugate momenta $\pi_m(\eta,x^i)=\dot{u}_m(\eta,x^i)$:
\bea
&& \left[u_0(\eta,{\bf x}),u_{0}(\eta,{\bf x}')\right] =
\left[\pi_0(\eta,{\bf x}),\pi_{0}(\eta,{\bf x}')\right] = 0 \nn \\
&& \left[u_m(\eta,{\bf x}),u_{m'}(\eta,{\bf x}')\right] =
\left[\pi_m(\eta,{\bf x}),\pi_{m'}(\eta,{\bf x}')\right] = 0 \nn \\
&& \left[u_0(\eta,{\bf x}),\pi_0(\eta,{\bf x}')\right] = i \d^d({\bf x}'-{\bf x}) \nn \\
&& \left[u_m(\eta,{\bf x}),\pi_{m'}(\eta,{\bf x}')\right] = i \d(m-m')\d^d({\bf x}-{\bf x}');
\label{fieldscomm}
\eea
we then express them in terms of their Fourier components:
\bea
u_m(\eta,x^i) &=& \int \frac{d^dk}{(2\pi)^{d/2}} \left[u_{m,k}(\eta) a_m(k) e^{i{\bf k}\cdot {\bf x}}
+ u^*_{m,k}(\eta) a^\dag_m(k) e^{-i{\bf k}\cdot {\bf x}} \right] \nn \\
\pi_m(\eta,x^i) &=& \int \frac{d^dk}{(2\pi)^{d/2}} \left[\dot{u}_{m,k}(\eta) a_m(k) e^{i{\bf k}\cdot {\bf x}}
+ \dot{u}^*_{m,k}(\eta) a^\dag_m(k) e^{-i{\bf k}\cdot {\bf x}} \right].
\label{fieldexp}
\eea
The operators $\{a_m(k)\}$ can be made to obey a canonical oscillator algebra
\bea
&& \left[ a_0({\bf k}),a_0({\bf k}') \right] =
\left[ a^\dag_0({\bf k}),a^\dag_0({\bf k}') \right] = 0 \nn \\
&& \left[ a_m({\bf k}),a_{m'}({\bf k}') \right] =
\left[ a^\dag_m({\bf k}),a^\dag_{m'}({\bf k}') \right] = 0 \nn \\
&& \left[ a_0({\bf k}),a^\dag_0({\bf k}') \right] = \d^d({\bf k}-{\bf k}') \nn \\
&& \left[ a_m({\bf k}),a^\dag_{m'}({\bf k}') \right] = \d(m-m')\d^d({\bf k}-{\bf k}'),
\label{acomm}
\eea
and therefore they can be interpreted as a set of creation--annihilation operators,
provided the wavefunctions $\{u_{m,k}(\eta)\}$ satisfy the equation
(which has, again, a Schr\"{o}dinger-like form):
\be
\ddot{u}_{m,k}(\eta)+  \omk^2(\eta) u_{m,k}(\eta)=0,
\label{modeequation}
\ee
where we have denoted
\be
\omk(\eta) = \sqrt{k^2+ m^2a^2(\eta) - \frac{\ddot{\xi}_m(\eta)}{\xi_m(\eta)}}=
\sqrt{k^2 - V_{m,k}(\eta)},
\label{defomega}
\ee
and they are normalized according to the wronskian condition (as expected)
\be
u_{m,k}(\eta)\dot{u}^*_{m,k}(\eta)- u^*_{m,k}(\eta)\dot{u}_{m,k}(\eta) = i.
\label{KGnorm}
\ee

We should stress that, even if the massive fields $u_m$ (with $m\neq0$) do not have the
correct canonical dimension, $[u_m]=[M^{\frac{d-2}{2}}]$, this discrepancy is compensated
by the dimension of the creation--annihilation operators, as can be deduced from (\ref{acomm}).
So the mode coefficients $u_{m,k}$ still have the proper canonical dimension of quantum-mechanical
wavefunctions: $[u_{m,k}]=[M^{-\frac{1}{2}}]$. This will be important when we will
need to normalize the initial fluctuations, as we will do in the next section.

\section{Derivation of the spectrum}
\label{specder}

The aim of this section is to calculate the spectral distribution for both the massless
and the massive gravitational modes, using the actions developed in the last section.
Hence, let us imagine a phase transition in which the Universe evolves from the inflationary
regime described by the Kasner solution presented in section \ref{backgroundsect} to a simple
final Minkowski era. Since we are interested in models that have 3 external dimensions, from now on we will
assume $d=3$. Moreover, to have an inflationary expanding initial phase, we need, as observed before,
2 or more internal dimensions. For the sake of simplicity we will take the value $n=2$, but
the consideration we will make can easily be extended to a different number of internal dimensions.
The phase transition will be analysed by making use of the sudden transition approximation (see,
for example, \cite{Maggiore:1999vm}), so that the geometry changes instantaneously at the conformal
time $-\eta_1$. This approximation gives reliable results only if the frequency of the amplified
modes is much lower than the transition velocity, which is estimated by the
curvature scale at the end of inflation $H_1 \sim 1/\eta_1a_1 =k_1/a_1 = k_1$ (of
course $a_1=a(-\eta_1)=1$). For this reason, $H_1$ represent a cut-off frequency.

In this framework, the Bogoliubov coefficients that describe the transformation from the
$|in \rangle$ to the $|out \rangle$ states can be obtained by simply imposing continuity
of the Fourier coefficients $u_{m,k}$ and their time derivative at the transition time $-\eta_1$.
This procedure is nevertheless effective only if we have an unambiguous way to normalize the $|in \rangle$
functions to pure positive norm states. This is indeed what happens in our case, since
it is easy to see that the background solutions, written in conformal time, are
\bea
a(\eta) = \left(-\frac{\eta}{\eta_1}\right)^{\lb},&&
b(\eta) = \left(-\frac{\eta}{\eta_1}\right)^{\mb} \nn \\
\lb=\frac{\l}{1-\l},&&\mb=\frac{\mu}{1-\l},
\label{confbacksol}
\eea
with the parameters $\l$ and $\mu$ given in (\ref{tsol}); since, if we want an inflationary solution,
it must be $\l <0$, we see that $\a(\eta)$ vanish as $\eta \ra - \infty$. On the other side, the expression
for the pump field during the Kasner regime is (note that it is independent from the number
of internal and external dimensions):
\be
\xi_0(\eta) = \sqrt{\frac{M^{d-1}}{2}}\frac{1}{\psi_0(0)}\sqrt{-\frac{\eta}{\eta_1}},~~~~~~
\xi_m(\eta) = \sqrt{\frac{M^d}{2}}\frac{1}{\psi_m(0)}\sqrt{-\frac{\eta}{\eta_1}},
\label{KasnerPF}
\ee
but of course the ratio $\ddot{\xi_m}/\xi_m$ always has the same  behaviour, like $\eta^{-2}$.
So the whole potential $V_{m,k}(\eta)$ goes to zero at the infinite past, and in this limit
the mode equation reduces to that of a massless non-interacting field, $\ddot{u}_{m,k}+k^2 u_{m,k} = 0$.
Moreover, as explained in section \ref{Actsection}, the dimension of the mode coefficient is the same
as expected, so we can guess the correct initial expression for it:
\be
u_{m,k}(\eta) = \frac{1}{\sqrt{2k}}e^{-i|k|\eta}.
\label{modeiniz}
\ee

With this in hand, what we need to do is to find a complete solution for
(\ref{modeequation}) fixing the free parameters in a suitable way to match the
asymptotic solution (\ref{modeiniz}), and then to solve the linear system
\bea
&& u_{m,k}^{(\rm in)}(-\eta_1) = \a_m(k) u_{m,k}^{+(\rm out)}(-\eta_1)+\b_m(k)
u_{m,k}^{-(\rm out)}(-\eta_1) \nn \\
&& \dot{u}_{m,k}^{(\rm in)}(-\eta_1) =
\a_m(k) \dot{u}_{m,k}^{+(\rm out)}(-\eta_1)+\b_m(k) \dot{u}_{m,k}^{-(\rm out)}(-\eta_1),
\label{Bogsystem}
\eea
to find the Bogoliubov coefficient $\b_m(k)$. Here $u_{m,k}^{+(\rm out)}$ and $u_{m,k}^{-(\rm out)}$
are respectively the positive- and negative-frequency eigenfunctions in the Minkowski phase.
The Bogoliubov coefficient represents the amount of particles created by the gravitational
fields in the mass interval $[m,m+dm]$. From this we can obtain the energy density per
logarithmic interval unit for each mode:
\bea
\frac{d\rho_0(k)}{d\log k} &=& \frac{k^4}{\pi^2} |\b_0(k)|^2  \label{0spectrexpr} \\
\frac{d\rho_m(k)}{d\log k} &=& \frac{k^3}{\pi^2}\sqrt{k^2+m^2}|\b_m(k)|^2 dm.
\label{KKspectrexpr}
\eea
These expressions should be normalized with their respective coupling constants
(\ref{4DPlankmass}) and (\ref{4Dmasscoupling}). Note that the different dimensions of the
spectral distributions are balanced by the different dimensions of the massless and massive
coupling constants, which explains the quantization procedure with the ``odd'' dimensions of section
\ref{Actsection}. Then the massive spectrum should be integrated over all masses (eventually
one can be interested in the contribution of a particular mass interval).
To obtain a dimensionless quantity, we then choose to normalize the spectral
distributions just obtained to the scale curvature $H_1$ at the end of inflation. The spectral
distribution can therefore be cast in its final form:
\be
\Om(k) = \Om_0(k)+\Om_{{ \rm KK}}(k),
\label{Totspec}
\ee
with
\bea
&& \Om_0(k)= \frac{k^4}{\pi^2 H_1^2 \Mp^2} |\b_0(k)|^2 \label{0finalspec} \\
&& \Om_{{ \rm KK}}(k)= \int dm~\Om_m(k) = \int dm~\frac{k^3}{\pi^2 H_1^2 M_m^3}\sqrt{k^2+m^2}|\b_m(k)|^2.
\label{KKfinalspec}
\eea
In the next sections we will apply this procedure to derive explicitly the spectral amplitudes.

\section{Spectral distribution for the massless mode}
\label{masslesssect}

The equation for the Fourier modes (\ref{modeequation}) becomes, in the massless case:
\bea
\ddot{u}_{0,k} + \left[ k^2 + \frac{1}{4\eta^2}\right] u_{0,k}&=&0~~~~~~~~~~~~~\eta<-\eta_1 \nn \\
\ddot{u}_{0,k} +  k^2 u_{0,k}&=&0 ~~~~~~~~~~~~~\eta>-\eta_1,
\label{0modeequations}
\eea
so the $|in \rangle$, solution which asymptotically tends to (\ref{modeiniz}), and the
$|out \rangle$ solutions are:
\bea
u_{0,k}^{(\rm in)}&=& \sqrt{\frac{\pi}{4}|\eta|}H_0^{(2)}(k\eta) \nn \\
u_{0,k}^{+(\rm out)}&=& \frac{e^{-ik\eta}}{\sqrt{2k}} \nn \\
u_{0,k}^{-(\rm out)}&=& \frac{e^{ik\eta}}{\sqrt{2k}},
\label{0modesolutions}
\eea
where $H_0^{(2)}$ is the Hankel function of the second kind.
This gives, for the Bogoliubov coefficient $\b$, the expression
\be
\b_0(k) = \sqrt{\frac{\pi}{32k\eta_1}} e^{ik \eta_1} \left[2k\eta_1 H_0^{(2)}(k\eta_1)
+ i \left(2H_0^{(2)'}(k\eta_1) k\eta_1 - H_0^{(2)}(k\eta_1)\right) \right].
\label{0bogcoeff}
\ee
The expression in square brackets can now be approximated for small values of the argument
of the Hankel functions, since our approximation is valid if $k\eta_1\ll1$, and the leading term
is logarithmic. So the energy density (\ref{0spectrexpr}) takes the form\footnote{In plotting the
spectrum we do not need to use the approximation. Actually, in the forthcoming calculations,
we heavily rely on numerical computations.}
\be
\frac{d\rho_0}{d\log k} = \frac{H_1^4}{8\pi^3}\left(\frac{k}{k_1}\right)^3\log^2\frac{k}{k_1}.
\label{0energyexpr}
\ee

\begin{figure}[h]
\begin{center}
\epsfig{file=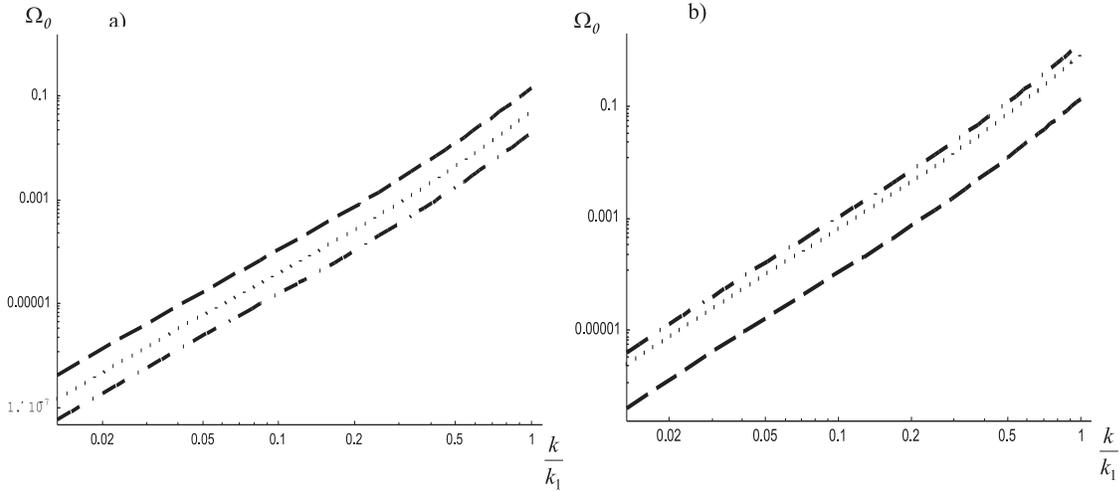,width=15cm}
\caption{The spectral amplitude of the massless mode. In a) the spectrum
is evaluated at $\a=-1$ and for $z_0H_1=7$ (dashed line), $z_0H_1=20$ (dotted line)
and $z_0H_1=50$ (dot-dashed line). In b) the spectrum
is evaluated at $z_0H_1=-7$ and for $\a=-1$ (dashed line), $\a=-5$ (dotted line)
and $\a=-8$ (dot-dashed line).}
\label{Spec0}
\end{center}
\end{figure}

The next step is to solve eq. (\ref{pertsep1}) for the massless mode. Its general solution can be
written as\footnote{We recall that, in what follows, the numerical values are only valid in $d=3$ and
$n=2$ dimensions.}
\be
\psi_0(z) = c_0 + c_1 \frac{z_0}{1-5\a} \left( 1+ \frac{z}{z_0} \right)^{1-5\a},
\label{zsolutionw}
\ee
but the normalization condition (\ref{zeronormcond}) imposes $c_1=0$, so the only acceptable
solution is the constant one $\psi_0(z) = c_0$. The free parameter $c_0$ can be calculated using
(\ref{zeronormcond}) as well, so we get:
\be
\psi_0 = \sqrt{-\frac{1+5\a}{2Mz_0}}.
\label{solpsi0}
\ee
Finally, the expression for (\ref{0finalspec}) is
\bea
&& \Om_0(k) = \frac{1}{8\pi^3}\left(\frac{H_1}{\Mp}\right)^2
\left(\frac{k}{k_1}\right)^3\log^2\frac{k}{k_1}, \label{0finalres} \\
&& \Mp^2 = \frac{z_0M^3}{2(-1-5\a)},
\label{0finalMp}
\eea
and it is plotted in Fig. \ref{Spec0} for different values of the parameters $z_0H_1$ and $\a$.

\section{The KK spectrum}
\label{KKsect}

First of all we will evaluate the massive coupling constant. In order to do
this we have to solve eq. (\ref{auxperteq}) for $m\neq0$. Using the background
solutions (\ref{solCLP}), it is easy to see that this equation is again a Bessel-like equation
of the form:
\be
\psih''_m + \left( m^2 - \frac{5\a(5\a-2)}{4} \frac{1}{(z+z_0)^2} \right)\psih_m = 0,
\label{zBessel}
\ee
and its general solution can be written as
\be
\psih_m(z) = \sqrt{z+z_0} \left[c_1(m) J_{\nu}(m(z+z_0))+c_2(m) Y_{\nu}(m(z+z_0))\right],
\label{zsolprovv}
\ee
where $J_\nu$ and $Y_\nu$ are of course the Bessel functions of the first and of the
second kind, and the parameter $\nu$ is related to the dilaton coupling constant by the
relation:
\be
\nu=\frac{1-5\a}{2}.
\label{nu}
\ee
The constants $c_1$ and $c_2$ are fixed by using the junction condition (\ref{Israelexp})
and the normalization condition (\ref{norm1}). From the junction condition we get,
using some algebraic properties of Bessel functions
\be
c_2=-\frac{J_{\nu-1}(mz_0)}{Y_{\nu-1}(mz_0)}c_1.
\label{ccond1}
\ee
From the second one, using their orthonormality relation\footnote{See for example the classical
electrodynamics textbook by J.D. Jackson},
we get
\be
c_1(m) = \sqrt{m}\frac{Y_{\nu-1}(mz_0)}{\sqrt{J_{\nu-1}^2(mz_0)+Y_{\nu-1}^2(mz_0)}}.
\label{ccond2}
\ee
Finally we can write the properly normalized solution of (\ref{zBessel}) as
\be
\psih_m(z) = \sqrt{m(z+z_0)} \frac{ \left[Y_{\nu-1}(mz_0)
J_{\nu}(m(z+z_0))- J_{\nu-1}(mz_0) Y_{\nu}(m(z+z_0))\right]}{\sqrt{J_{\nu-1}^2(mz_0)+Y_{\nu-1}^2(mz_0)}}.
\label{msol}
\ee

\begin{figure}[h]
\begin{center}
\epsfig{file=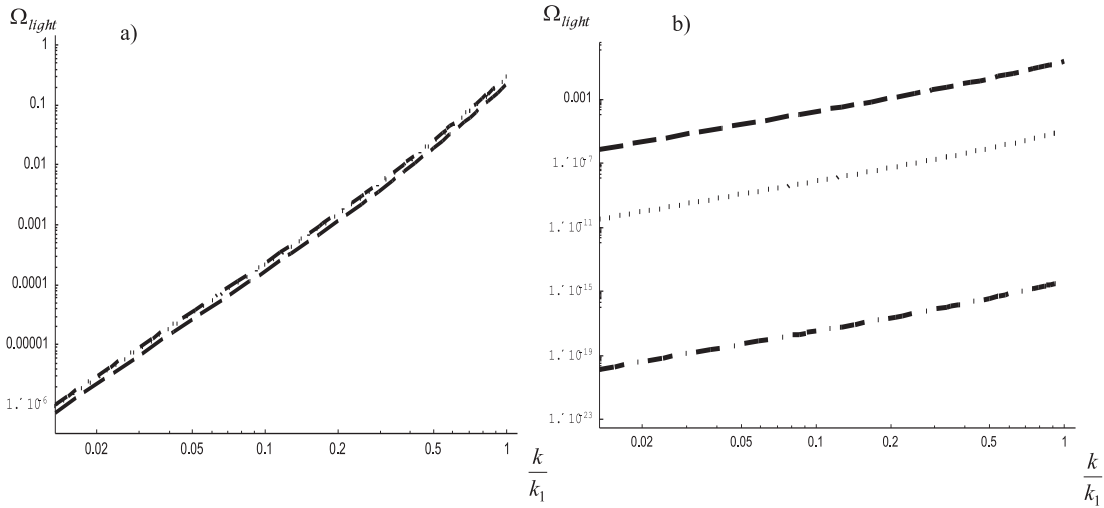,width=15cm}
\caption{The spectral amplitude of light KK modes. As before, in a) the spectrum
is evaluated at $\a=-1$ and for $z_0H_1=7$ (dashed line), $z_0H_1=20$ (dotted line)
and $z_0H_1=50$ (dot-dashed line). The three curves overlap, so that they are quite confused.
In b) the spectrum is evaluated at $z_0H_1=-7$ and for $\a=-1$ (dashed line), $\a=-5$ (dotted line)
and $\a=-8$ (dot-dashed line).}
\label{KKlight}
\end{center}
\end{figure}

Now we turn to consider eq. (\ref{modeequation}). Actually it is very hard to solve it in its complete form,
so the best we can do is to seek for particular limits in which the equation simplifies a little.
As already pointed out, in the phase transition we considered, an energy scale naturally emerges, i.e.
the curvature scale at the end of the inflationary epoch. In order to have a complete
understanding of the effects of the KK modes, we will study
what happens to lighter or heavier modes as regards to this curvature scale.
Firstly, we limit our attention to modes that are lighter than
the curvature scale at the end of inflation, $m \ll H_1$. In this case we can neglect
the mass term in (\ref{defomega}), so that the mode equation reduces to the same as
obtained for the massless mode, eq. (\ref{0modeequations}). It is easy to see
that even if the $|out \rangle$ solutions are massive waves, the spectral distribution
remains unchanged, as in eq. (\ref{0energyexpr}). So, the only contribution to the integral, which must
be calculated from $m=0$ to $m=H_1$, comes from the massive coupling constant.
The integration leads to a change of the mass parameter that controls the
normalization of the spectral amplitude, while the shape of the spectrum remains
unchanged. We obtain
\bea
&& \Om_{{\rm light}}(k) = \int_0^{H_1} dm~\Om_m(k) =
\frac{1}{2\pi^3}\left(\frac{H_1}{M_*}\right)^2
\left(\frac{k}{k_1}\right)^3\log^2\frac{k}{k_1}, \label{lightfinalres} \\
&& M_*^2 = \left( \int_0^{H_1}\frac{dm}{M_m^3} \right)^{-1}
= \frac{M^3}{4}\left(\int_0^{H_1}dm~|\psi_m(0)|^2\right)^{-1}.
\label{lightfinalM}
\eea
In Fig. \ref{KKlight} we show a numerical estimate of the behaviour of the spectrum for different
values of the parameters.

\begin{figure}[h]
\begin{center}
\epsfig{file=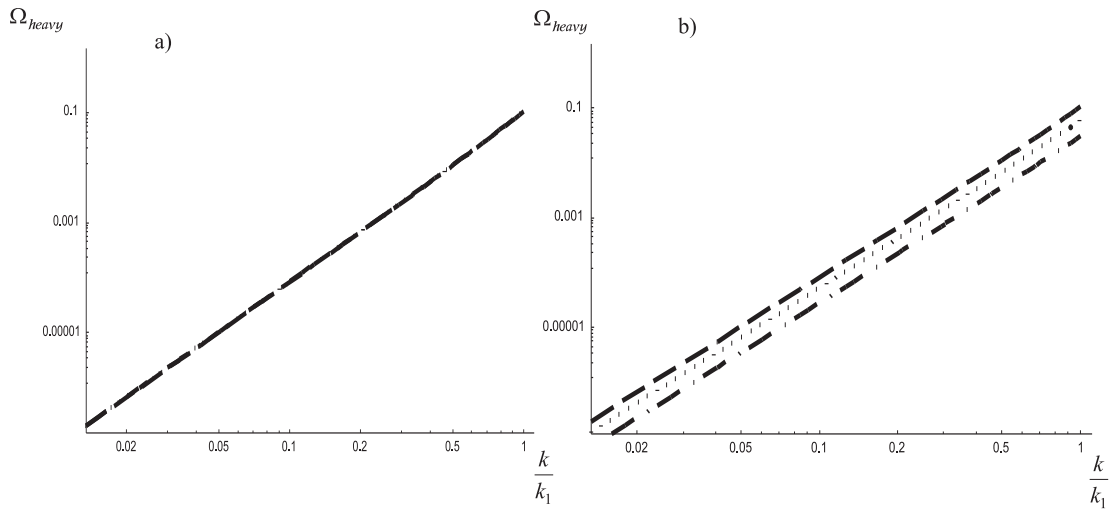,width=15cm}
\caption{The spectral amplitude of the heavy KK modes. As in the other figures, plot a)
represents the spectrum evaluated at $\a=-1$ and for $z_0H_1=7$ (dashed line),
$z_0H_1=20$ (dotted line) and $z_0H_1=50$ (dot-dashed line). Again, overlapping of the curves
do not permit to distinguish them. Plot b) represents the spectrum
evaluated at $z_0H_1=-7$ and for $\a=-1$ (dashed line), $\a=-5$ (dotted line)
and $\a=-8$ (dot-dashed line)}
\label{KKheavy}
\end{center}
\end{figure}

Next, we try to solve (\ref{modeequation}) for the heavy modes.
To do this, we use a WKB-like approximation for the mode function
$u_{m,k}$ \cite{Audretsch:1979uv,Lawrence:1995ct}, which can be written as:
\be
u_{m,k}^{(in)} \simeq \frac{1}{\sqrt{2\omk(\eta)}} \exp
\left[-i \int^{\eta} d\eta' \omk(\eta') \right].
\label{WKBmode}
\ee
This approximation is valid if the variation of the frequency is small with respect to the
frequency itself. More precisely, (see, for example, \cite{Birrell:1982ix}), one must have
\be
\e = \frac{3}{4} \frac{\dot{\om}_{m,k}^2}{\omk^4} - \frac{1}{2}\frac{\ddot{\om}_{m,k}}{\omk^3} \ll 1.
\label{epspar}
\ee
Since the parameter $\e$ grows monotonically in time, we only need to check where our approximation
is valid at the transition epoch $\eta_1$, and we are guaranteed that it
always holds before that time\footnote{To be more precise, $\e$ actually reaches a maximum,
but well after the transition time.}. It is not difficult to see that
$|\e(-\eta_1)|\ll 1$ if $ma_1\gg k_1$, which is exactly the regime we wish to investigate.
This is not an unexpected result, since it is known
\cite{Birrell:1982ix} that this approximation works better if the transition
is adiabatic, so that the variation of the curvature is slow with respect to the energy.
But this means that the sudden transition approximation is no longer valid. Nevertheless,
we could still get clues on the form of the spectrum by analytically continuing the function
$\omk$ and evaluating the integral in (\ref{WKBmode}) on a suitable path.
We consider a semicircle on the upper half-plane. Its radius should be chosen to be greater than
$\eta_1$, to stay in the region in which the WKB approximation is valid, but not too big,
since we expect the mass term to be dominant with respect to the pure frequency
term $k^2$. In this regime the function $\omk$ can be approximated as
\be
\omk(\eta) \simeq m a(\eta) + \frac{k^2}{2m a(\eta)} - \frac{1}{2m a(\eta)}\frac{\ddot{\xi}_m}{\xi_m},
\label{approxomega}
\ee
and has, as one could expect, the form of non-relativistic energy for a massive particle,
plus a term that describes the direct interaction between the particle and the background geometry,
which however is negligible.
Note that, in the approximation (\ref{approxomega}), $\omk$ has only a singularity at the origin, so
we can actually shrink the radius of the integration path till $R \simeq \eta_1$ without
changing the result of the integration \cite{Lawrence:1995ct}.
It is not difficult to see that the square modulus of the Bogoliubov coefficient is given by:
\be
|\b_m(k)|^2 = \exp\left[2~{\rm Im}\left(\int d\eta~\omk(\eta)\right) \right].
\label{bogteor}
\ee
The integral is easy to solve if we make use of the approximation (\ref{approxomega}), so we get:
\be
|\b_m(k)|^2 = \exp\left[2\sin \pi \lb \left( \frac{m a_1}{k_1(1+\lb)}-\frac{k^2}{2 m a_1 k_1(1-\lb)}
+\frac{k_1}{8ma_1(1+\lb)} \right) \right],~~~~~~m \gg H_1.
\label{KKbogcoeff}
\ee
This expression can now be substituted in (\ref{KKfinalspec}) to obtain the final form for the
Kaluza--Klein spectrum. The integral in this case runs from $m=H_1$ to infinity.
\be
\Om_{\rm heavy}(k)= \int_{H_1}^{+\infty} dm~\frac{k^3}{\pi^2 H_1^2 M_m^3}\sqrt{k^2+m^2}|\b_m(k)|^2.
\label{SpecHeavy}
\ee
Of course this integration can only be carried out numerically. The result is presented in
Fig. \ref{KKheavy} for the same values of the parameters $z_0H_1$ and $\a$ used in the
previous computations.

\section{Comments and conclusions}
\label{conclsec}

In this paper we have presented a simple model in which a brane coupled with
a bulk dilaton evolves from an initial inflationary Kasner phase to a final Minkowski era.
Then we have studied the evolution of the tensor perturbations on this background, paying
particular attention to their normalization to an initial state of vacuum fluctuation.
We have found that the massless and the massive modes can be treated independently, and
we have evaluated (analytically where it was possible, numerically elsewhere) the spectrum
of the tensor perturbations for the massless mode (to be identified with the graviton)
and for both the ultrarelativistic and the non-relativistic massive modes. Of course the
total spectrum an observer would detect should be approximatively the sum of the three.

\begin{figure}[h]
\begin{center}
\epsfig{file=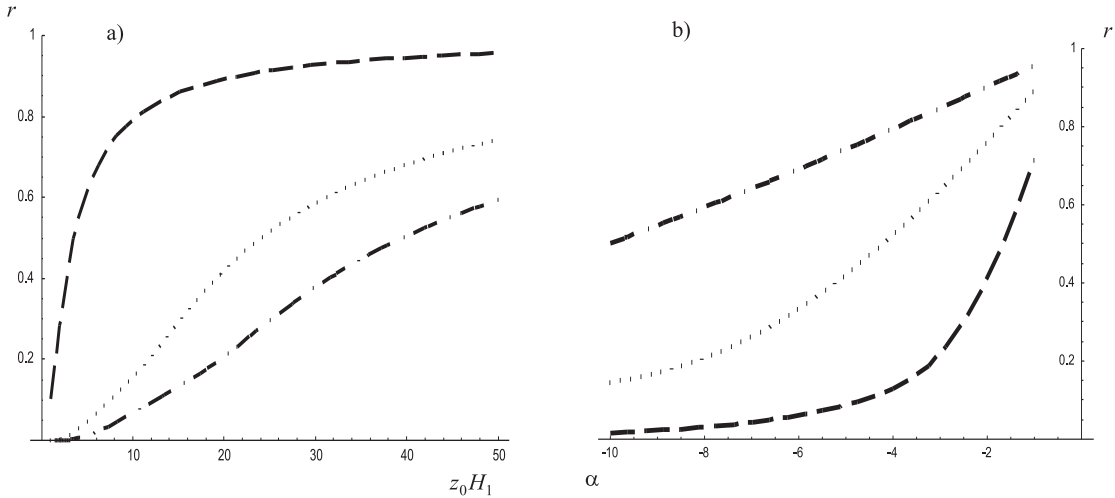,width=15cm}
\caption{Plot a) shows the behaviour of $r$ with respect to $z_0H_1$ for the three values
of $\a$ used elsewhere: $\a=-1$ (dashed line), $\a=-5$ (dotted line)
and $\a=-8$ (dot-dashed line). Plot b) is complementary to the first one, and shows the
behaviour of $r$ with respect to $\a$ for the three different values of $z_0H_1=7$ (dashed line),
$z_0H_1=20$ (dotted line) and $z_0H_1=50$ (dot-dashed line).}
\label{plotr}
\end{center}
\end{figure}

The behaviour of the plot we presented suggests some comments: As widely expected, the
relative importance of the KK corrections on the total spectrum grows as the AdS
curvature increase\footnote{Note that we have chosen to normalize
the spectrum to the end of inflation curvature, so that an increase of the AdS curvature
actually corresponds to an increase of the curvature scale at which the transition occurs. This is
the reason why it seems that the massless mode spectrum is influenced by the change in
$z_0$ and the KK spectrum is not.},
because \cite{Shiromizu:1999wj,Koyama:2000cc,Langlois:2000ns} deviations from classical general
relativity become more and more relevant as the energy increase. On the other hand,
the coupling between the dilaton and the brane has the opposite effect. In fact,
as $\s_1$ approaches its limiting value $2/\sqrt{p}$ (see eq. (\ref{diseq})), the massive
perturbations are highly suppressed. We can have a better understanding of the relative importance
of the contribution of the KK modes by defining the ratio:
\be
r=\frac{\Om_{\rm KK}}{\Om_0+\Om_{\rm KK}},
\label{r}
\ee
and by plotting its behaviour with respect to the two parameters we are interested in. This
is done in Fig. \ref{plotr}.

As we can see, the KK contribution is very low at small values of $z_0H_1$; however, it rapidly increases,
to become completely dominant in the high energy regime, differently from what happens in other
models present in the literature \cite{Langlois:2000ns,Gorbunov:2001ge,Kobayashi:2003cn,Kobayashi:2005dd}.
Of course, this effect could in principle be relevant in observational tests concerning the amplification of
tensor perturbation during inflation. Actually, since the spectrum obtained in our simple evolution model
increases with the frequency, it is not difficult to tune the free parameters so as to
satisfy COBE and pulsar constraints. Moreover, the relevance of the massive modes contribution can be lowered
by the presence of the dilaton. In fact, at relatively high energies, the effect of the bulk gravitons can be
very strong if the dilaton is absent or weakly coupled, but it becomes negligible as the coupling increases.
This indicates that, even at high energies, the enhancement of the amplitude of the KK spectrum
can be ``cured'' by a suitable choice of coupling parameter, leading to a spectrum that is quite similar
to the 4-dimensional one.

The decisive test for models with strong production of KK modes will be the future detection
of stochastic gravitational waves at gravitational antennae. For this purpose, more accurate and realistic
models are needed, in which the presence of matter on the brane is taken into account, and a smooth transition
from the inflationary to the standard cosmological evolution is considered. This will be the
subject of forthcoming papers to investigate which features of the simple model under discussion
can be generalized to these more realistic models.

\section*{Acknowledgements}

It is a pleasure to thank Marco Ruggieri for hints about numerical integration,
and Maurizio Gasperini and Gabriele Veneziano for helpful discussions and comments on the paper.

\end{document}